\documentclass[10pt,pre,superscriptaddress,twocolumn,showpacs,floatfix,noeprint]{revtex4-1}
\usepackage[utf8]{inputenc} % Para usar com codificacao de caracter utf8
\usepackage{mathtools}
\usepackage{amsfonts}
\usepackage{amsmath}
\usepackage{amsthm}
\usepackage{color}
\usepackage{graphicx}
\usepackage{multirow}
\usepackage{rotating}
\begin{document}

\title{On the order of the phase transition in the spin-$1$ Baxter-Wu model}

\author{L. N. Jorge}
\affiliation{Instituto Federal do Mato Grosso - Campus C\'{a}ceres,
Av. dos Ramires s/n, 78200-000, C\'{a}ceres, MT, Brazil}
\author{L. S. Ferreira}
\affiliation{Universidade Federal de Goi\'{a}s, Av. Esperan\c{c}a s/n, 74.690-900, Goi\^{a}nia, GO, Brazil}
\author{A. A. Caparica}
\affiliation{Universidade Federal de Goi\'{a}s, Av. Esperan\c{c}a s/n, 74.690-900, Goi\^{a}nia, GO, Brazil}
\email{caparica@ufg.br}

\begin{abstract}
In this work we investigate the order of the phase transition of the spin-1 Baxter-Wu model. We used extensive 
entropic simulations to describe the behavior of quantities which reveal the order of the phase
transition. We applyied finite-sizing scaling laws for continuous and discontinuous phase transitions. Our
results show that this system exhibits an indeterminacy regarding the order of the phase transition, i.e., 
the results are conclusive for both transitions, whether continuous or discontinuous. In such a scenario we
carried out a study of the configurations in the region of the phase transition, which confirmed that the model
seems to undergo a tetracritical transition, with the coexistence of a ferromagnetic and three ferrimagnetic 
configurations, suggesting that it may be a multicritical point belonging to a critical line of 
an external or a crystalline fields, where the continuous and the discontinuous phase transitions may
coexist reflecting different features of the system. 
 \end{abstract}

\maketitle

\section{Introduction}

The order-disorder transitions in two-dimensional systems have been 
studied experimentally in structures constituted by adsorbed atoms or
molecules in single crystals and their universality class has been
experimentally determined. Among them there are some real systems with 
triplet interactions which belong to the same universality class of
the $q=4$ Potts model. Examples of such compounds are the chemisorbed overlayer 
$p(2\times 2)$ oxygen on Ni$(111)$\cite{Roelofs1981}, the 
adsorption system O/Ru$(0001)$\cite{Piercy1987} at $1/4$ 
monolayer and the $(2\times 2)$-$2H$ structure on Ni($111$)
\cite{Schwenger1994}. They are well-described by a model that 
considers three spin interactions in a triangular lattice.

In 1972 D. W. Woods and H. P. Griffiths\cite{Wood1972} proposed such a model:
the Baxter-Wu model(BW). They propounded a system of 
spins defined in a two-dimensional triangular lattice, with the spins 
located at the vertices of the triangles, and assuming the integer 
values $\pm 1$. 
An interesting feature of this spin system is that it exhibits an order-disorder 
transition, but it is not symmetric by inversion of all spins.
This model, known as the spin-$1/2$ case, was exactly solved by R.J. Baxter and F.Y. Wu
\cite{Baxter1973, Baxter1974, Baxter1974a} yielding the same critical temperature of the 
Ising model on a square lattice $k_BT_c/J=2/ln(1+\sqrt{2})=2.2691853...$, and
critical exponents belonging to the $q=4$ Potts model universality class,
namely $\alpha=\nu=2/3$,  $\beta=1/12$, and $\gamma=7/6$. Simulations using the 
short-time scaling formalism corroborate the exact results for $\nu$ and 
$\beta$\cite{Santos2001}, although not within the error bars. The same occurs 
in entropic simulations of the $q=4$ Potts model\cite{Caparica2015b}.

This behavior of the BW model should be expected, since both models have 
the same symmetry and degree of degeneracy in the ground state. However, 
unlike the BW model, the $q=4$ Potts model demands logarithmic corrections.
Thereby, to investigate this apparent contradiction, Kinzel \textit{et al}
\cite{Kinzel1981} studied a generalization of the BW model by insertion 
of annealed vacancies. In this new scenario we have the spin-1 Baxter-Wu model 
where the spins variables can assume one of three discrete values $0$ and $\pm 1$. 
They found out through the behavior of the correlation length and by using finite 
size scaling techniques, that a second order transition would happen only for 
the pure BW model.

Notwithstanding, another study of the spin-1 BW model was
carried out by Costa and Plascak\cite{Costa2004}, using Monte Carlo 
simulations with the standard Metropolis algorithm. Applying finite size scaling 
considerations they obtained a critical behavior different from the first work, 
with a second order transition and well defined critical exponents.

In view of these conflicting resuls, our expectation in carrying out a
detailed study of the model was to detect a pure discontinuous behavior, as 
predicted by Kinzel or a continuous behavior, as obtained by Costa and Plascak.
Nevertheless our findings show that it exhibits a mixed behavior, typical of
multicritical points. 

A multicritical transition occurs in a region of the phase diagram where multiple
phases co-exist. In general, multicritical states emerge in the intersection of a
first order and a second order transitions. In such regions it is common the 
superposition of both transitions\cite{Riedel1972,Griffiths1973,Graf1967,Landau1971,Garland1971,Saul1974}.
Musial \cite{Musial2004} applyied the Monte Carlo method to the 3D Ashkin-Teller model in the region of
the three-critical point and found out an indeterminacy regarding the order of transition of the system.
Critical exponents point out to a continuous transition, while the Binder cumulant of energy evidences 
a discontinuous transition. Another characteristic observed near the tricritical point is the presence of 
a non-zero valley in the energy probability\cite{Simenas2014,Fytas2011a}. In a recent work 
Dias \textit{et. al.}\cite{Dias2017} carried out a study of the spin-1 Baxter-Wu model with a 
crystalline field and detected the existence of a pentacritical point with the coexistence of three ferrimagnetic
and a ferromagnetic configurations, along with that with spins zero. 

In the present work we are interested in the characteristics of the order of the phase
transition in the spin-$1$ Baxter-Wu model. Here we employ entropic sampling 
simulations to construct the density of states, which is estimated for systems with
non-multiple of three lattice sizes, following the same implementation of \cite{Jorge2016}. 
Some quantities which reveal the order of the transition, as the total magnetization, the 
energy, and the susceptibility are investigated.

This paper is organized as follows: Section II presents the BW model. 
In Sec. III we provide details of our simulations. The finite size scaling analysis is 
showed in section IV. The results are discussed in section V, and in section VI a summary 
and outlooks are given.

\section{The Baxter-Wu Model}

In the Baxter-Wu model the three spin interaction is governed by the Hamiltonian 
\begin{equation}
 H_{BW}=-J\sum_{<i,j,k>}s_is_js_k, \label{eq.baxterwu}
\end{equation}
where the spin variables are located at the vertices of the lattice and take on the values $s_i=\pm 1$,$0$, $J$ 
plays the role of a coupling constant that fixes the energy scale, 
and the sum extends over all triangular faces of the lattice. 

\begin{figure}[ht]
\begin{center}
\includegraphics[width=1.\linewidth]{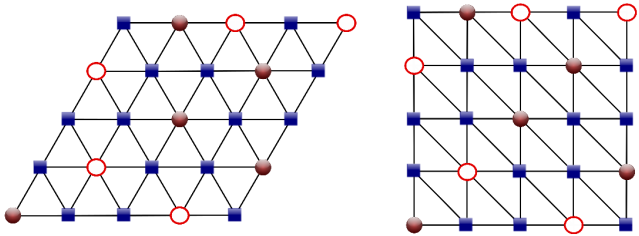}
\end{center}
% \vspace{-.5cm}
\caption{(color online). Left: Representation of a possible configuration of the Baxter-Wu model as 
a superposition of three sub-lattices on a triangular lattice.  Right:
Transposition of the triangular lattice to a square lattice. The cycles
represent spins +1, the squares spins -1, and the open cycles the zeros.
\label{baxterwu}} 
\end{figure}

In our work we deal with the system by transposing the triangular 
lattice to a square one, as shown in the right side of Fig.\ref{baxterwu}. 
One can see that each spin $\sigma_i\equiv s_{i,j}$  
has six triangular faces around it, and, if one visits all spins of 
the configuration computing the six triangular faces for each spin, 
each triangular face will be counted three times. Thus, the energy of
a particular configuration of lattice size $L$ defined by the 
Hamiltonian \eqref{eq.baxterwu} can be calculated in the square 
lattice as
%\begin{equation}\label{energy}
 \begin{align}
E =  &\frac{J}{3} \sum_{i=1}^{L}\sum_{j=1}^{L}s_{i,j}(s_{i-1,j}s_{i,j-1}+s_{i,j-1}s_{i+1,j-1} \notag \\
     &+s_{i+1,j-1}s_{i+1,j}+s_{i+1,j}s_{i,j+1}+s_{i,j+1}s_{i-1,j+1} \notag \\ 
     &+s_{i-1,j+1}s_{i-1,j}).  \label{energy}
 \end{align}
%\end{equation}

Likewise the $q=4$ Potts model and the spin-$1/2$ BW model, in this model, if one assumes multiple of three lattices,
four distinct ground states are present, one with all spins positive, giving a ferromagnetic configuration, and other 
three, where one sublattice remains positive while the other two are negated, yielding ferrimagnetic configurations
successively. However, if we choose non-multiple of three lattices, only the ferromagnetic configuration will
be present in the ground state and we can choose the order parameter of the system as the sum of all spins of the 
lattice
\begin{equation}
 M=\sum_{i,j=1}^{L}s_{i,j}.
\end{equation}
As shown in Ref. \cite{Jorge2016} the use of multiple or non-multiple of three lattices does not change the results
for the critical exponents, such that in the present study we adopted non-multiple of three lattices and the 
sum over all spins as the order parameter. In order to investigate the order of the phase transition of the 
model we performed entropic sampling simulations either treating the transition as discontinuous, either 
as continuous.

\section{Finite Size Scaling}

\subsection{Continuous phase transition}

From the fluctuations in the energy and the magnetization we can calculate the specific 
heat and the magnetic susceptibility
\begin{equation}
 C(T)=(\langle E^2 \rangle-\langle E \rangle^2)/T^2,
\end{equation}
\begin{equation}
 \chi(T)=(\langle m^2\rangle - \langle m \rangle ^2)/T,
\end{equation}
where $E$ is the energy, $m$ is the magnetization per site and $T$ is the absolute temperature.

Using the free energy definition and the appropriate derivatives of the free-energy density one can 
obtain scaling relations for various thermodynamic quantities\cite{Fisher1971,Fisher1972}, such as

\begin{equation}\label{eq:mag}
m_L(t)\approx L^{-{\beta}/\nu}\mathcal{M}(tL^{1/{\nu}}),
\end{equation}

\begin{equation}\label{eq:chi}
\chi_L(t)\approx L^{\gamma/\nu}\mathcal{X}(L^{1/\nu}t),
\end{equation}

\begin{equation}
c_L(t)\approx L^{\alpha/{\nu}}\mathcal{C}(tL^{1/{\nu}}),
\end{equation}
where $L$ is the linear size of the lattice, $t=(T_c-T)/T$ is the reduced temperature and $\alpha$, $\beta$,
$\gamma$ and $\nu$ are the static critical exponents that obey the scaling relation
\begin{equation}
 2-\alpha = d\nu=2\beta+\gamma,
\end{equation}
where $d$ is the dimensionality of the system. 
As $L\to \infty$, the pseudocritical temperature of the finite lattice obeys the scaling law
\begin{equation}\label{eq:tc}
 T_L=T_c+\lambda L^{-1/\nu}
\end{equation}
where $\lambda$ is a constant, $T_c$ is the critical temperature of the infinite system, and $T_L$ 
is the effective transition temperature for the lattice of linear size $L$.

One can obtain the critical exponent $\nu$ using the logarithmic derivatives of the magnetization
\cite{Ferrenberg1991,Caparica2000}.
\begin{equation}
[m^n]\equiv ln\frac{\partial\langle m^n\rangle}{\partial T},
\end{equation}
which have the same scaling behavior of the $4^{th}$ order Binder cumulant. Following the prescriptions of the 
works\cite{Caparica2000,Jorge2016},
we use the set of thermodynamic quantities that are functions of the logarithmic derivatives of the magnetization
\begin{equation}
 V_1\equiv 4[m^3]-3[m^4], \label{eq:V1}
\end{equation}
\begin{equation}
 V_2\equiv 2[m^2]-[m^4], \label{eq:V2}
\end{equation}
\begin{equation}
 V_3\equiv 3[m^2]-2[m^3], \label{eq:V3}
\end{equation}
\begin{equation}
 V_4\equiv (4[m]-[m^4])/3, \label{eq:V4}
\end{equation}
\begin{equation}
 V_5\equiv (3[m]-[m^3])/2, \label{eq:V5}
\end{equation}
\begin{equation}
 V_6\equiv 2[m]-[m^2]. \label{eq:V6}
\end{equation}
Using Eq. (\ref{eq:mag}) it is immediate to show that
\begin{equation}
 V_j\approx \frac{1}{\nu}\ln L + \mathcal{V}_j(tL^{\frac{1}{\nu}}), \label{eq:Vj}
\end{equation}
for $j=1,2,...,6$. At the critical temperature $T_c(t=0)$ the $\mathcal{V}_j$ are constants independent of the 
lattice size $L$, therefore we can estimate $1/\nu$  by the slopes of $V_j$ calculated at $T_L$.
With the exponent $\nu$ already obtained, the next step is calculating the critical temperature using Eq. \ref{eq:tc} 
and the exponents $\beta$ and $\gamma$ from the slopes of the log-log plot of Eqs. \ref{eq:mag} 
and \ref{eq:chi} at the critical temperature $T_c$. A model with continuous phase transition displays well defined
critical exponents and allows the determination of the critical temperature by the scaling relation of Eq. (\ref{eq:tc}).

\subsection{Discontinuous phase transition}

In a discontinuous phase transition the scaling relations defined in the continuous transition do not hold. Usually the 
thermodynamic properties, such as magnetization and specific heat scale with the system dimension.The critical 
temperature of the system may be determined using the cumulant of the energy that scales with the lattice size \cite{Challa1986}. 
The fourth order cumulant of the energy is defined as

\begin{equation}
 U_E(T)\equiv 1-\frac{\langle E^4\rangle_{T}}{3\langle E^2\rangle^{2}_{T}}.
\end{equation}
This function has a minimum \cite{Binder2010a,Vollmayr1993,Binder1997}, which temperature scales as
\begin{equation}\label{eq:tc_first}
 T_L=T_c+\lambda L^{-d}.
\end{equation}
where $d$ is the dimension of the system. Extrapolation for $L\rightarrow\infty$ gives the critical temperature 
$T_c$ for the infinite system.

The behavior of the cumulant of the magnetization, given by
\begin{equation}
 U_M(T)\equiv 1-\frac{\langle M^4\rangle_{T}}{3\langle M^2\rangle^{2}_{T}}.
\end{equation}
also may suggest a discontinuous phase transition. In this case we have the presence of a sharp minimum
and the crossing of the curves of different lattice sizes taking place around the critical temperature 
\cite{Lee1991,Challa1986}.  

Another way of finding the critical temperature is the analysis of the peaks of the energy probability.
In the region of the critical temperature this probability should have a double peak of same height and
the region between the peaks should have a null probability \cite{Lee1991,Binder1984,Binder1987}. 
In the canonical ensemble the energy probability is calculated as
\begin{equation}
P(E,T)= g(E) e^{-\frac{E}{k_BT}}.
\end{equation}
This method allows calculating the latent heat of the transition as the energy difference $\Delta E_L$ between
the two peaks, which obeys the scaling relation of $L^{-1}$ \cite{Lee1991}. The temperatures where the peaks reach the same
height also scales as Eq. \ref{eq:tc_first}.

If the entropy is given as a function of the energy one can calculate the microcanonical temperature directly
by the definition 
\begin{equation}
 \beta=\frac{1}{T}=\frac{\partial S}{\partial E},
\end{equation}
and the behavior of this quantity changes with the order of the phase transition. In the discontinuous 
transition it has a $s$-like shape and a straight line at the critical temperature divides it into two 
areas of same size\cite{Landau2014,Schnabel2011}.

\section{Entropic Simulations}

Our computational approach follows the Wang-Landau method \cite{Wang2001, Wang2001a}, except that 
we include some improvements proposed in \cite{Caparica2014, Caparica2012a, Ferreira2012a, Ferreira2012}, 
which are: i) the density of states is not updated at every spin flip, but, only after each Monte Carlo 
sweep, in order to pick uncorrelated configurations when we construct the density of states; 
ii) to avoid unnecessary extensive simulations, we perform them until $\ln f=\ln f_{final}$, defined by 
the canonical averages along the simulation. In \cite{Caparica2014} a checking parameter $\varepsilon$
was proposed. It signalizes automatically when the simulations should be halted. One consequence of 
adopting the checking parameter is that $f_{final}$ may be different for different runs; and, iii) at the 
beginning of the simulations, the microcanonical averages should not be accumulated before $\ln f=\ln f_{micro}$
(also defined by canonical averages during simulations), since, at the beginning of the simulations the 
configurations do not match those to maximum entropy. These three adjustments provide more accurate results, 
save CPU time and have the advantage of being easily implementable. 

Once the density of states is constructed, one can calculate the canonical average of any thermodynamic quantity $X$ as
\begin{equation}
 \langle X \rangle_T = \frac{\sum_E \langle X \rangle_E g(E) e^{-\beta E} }{\sum_E g(E) e^{-\beta E}},
\end{equation}
where $\langle X \rangle _E$ is the microcanonical average accumulated along the simulations, $\beta=1/k_BT$, where $T$ is 
the absolute temperature given in units of $J/k_B$ and $k_B$ is the Boltzman's constant.
\begin{figure} [b]
\centering
\hspace{2.0in}
\includegraphics[scale=0.34,angle=-90]{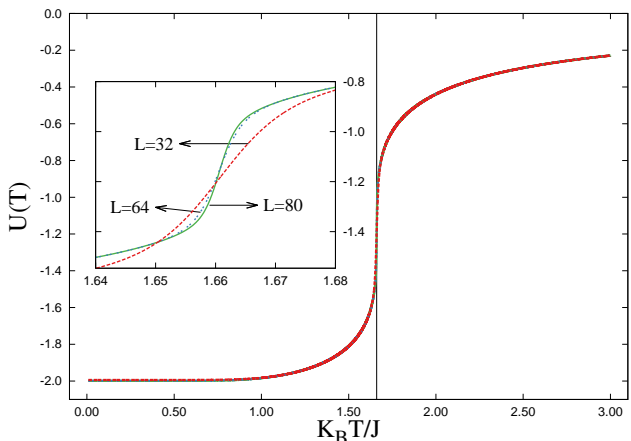}
\vspace{-0.20in}
\caption{Mean Energy. \label{fig:u_mean_80}}
\end{figure}
\begin{figure} [b]
\centering
\hspace{2.0in}
\includegraphics[scale=0.34,angle=-90]{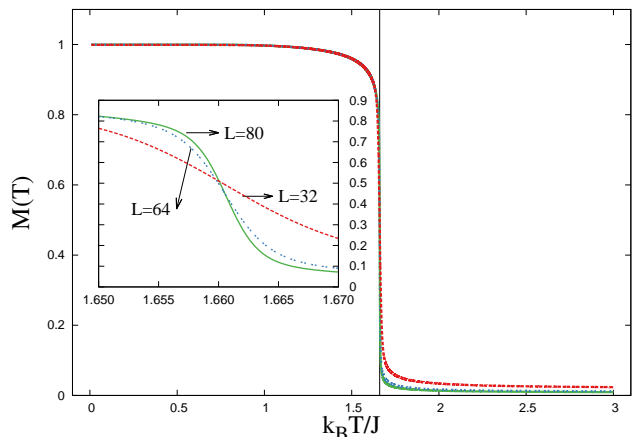}
\vspace{-0.20in}
\caption{Magnetization. \label{fig:mag_80}}
\end{figure}

\begin{figure} [h]
\centering
\hspace{2.0in}
\vspace{0.1in}
\includegraphics[scale=0.34,angle=-90]{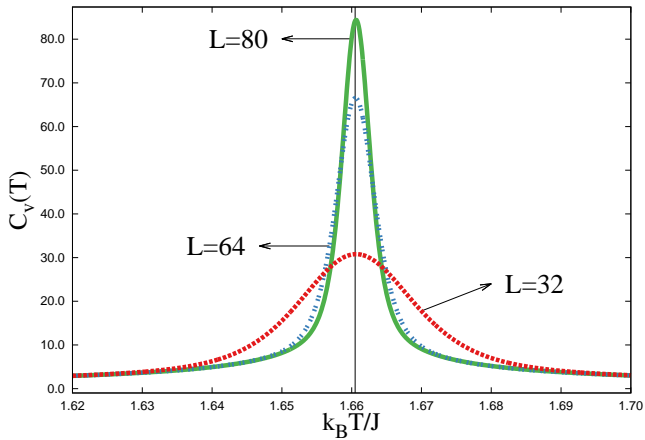}
\vspace{-0.20in}
\caption{Specific Heat of lattice sizes non multiple of three. \label{fig:cv}}
\end{figure}

\begin{figure} [h]
\centering
\hspace{2.0in}
\includegraphics[scale=0.34,angle=-90]{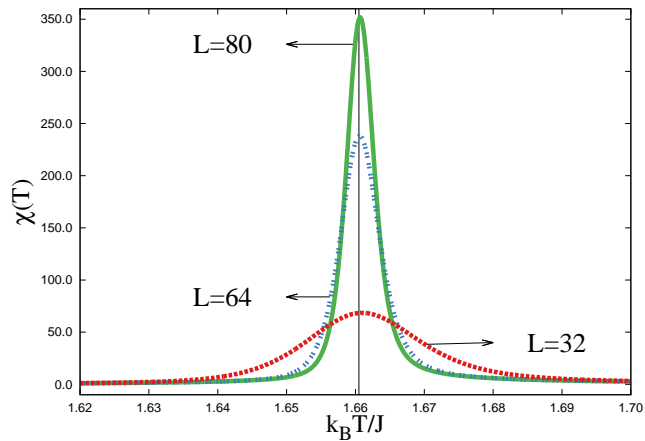}
\vspace{-0.20in}
\caption{Susceptibility of lattice sizes non multiple of three. \label{fig:khi}}
\end{figure} 

In Ref.\cite{Caparica2014} it was also noted that two independent similar finite-size scaling procedures can lead to
very different results for the critical temperature and exponents, which often do not agree within 
the error bars. To overcome this difficulty we carry out 10 independent sets of finite-size scaling procedures. 
The final results for the critical exponents and critical temperature
are obtained as an average over all sets.

\section{Results}

We study the spin-$1$ BW model on a square lattice with $N=L\times L$ sites, by performing the entropic simulations 
described above. We perform ten studies of finite
size scaling using the following lattice sizes : $L=$ 32, 40, 44, 52, 56, 64, 76, 80, 86, and 92, with
$N=$ 24, 20, 20, 16, 16, 16, 12, 12, 12, and 12 runs for each size, respectively. The choice of theses lattice sizes was based 
in the work of Jorge \textit{et al} \cite{Jorge2016}, where it was shown that one can simulate the BW model using 
non-multiple of three lattices and obtain the same results of the multiple of three ones.

Before presenting the results for continuous and discontinuous phase transitions, we analyze the behavior of energy, 
magnetization, specific heat, and magnetic susceptibility. Figs. \ref{fig:u_mean_80} and \ref{fig:mag_80} show the 
energy and the magnetization as functions of temperature for $L=32,64,$ and $80$. The main graphs suggest a discontinuous
phase transition with an abrupt variation in the region of the critical temperature. But if we look at the zoom in the
insets in a narrow range of temperature, we observe a more likely continuous behavior. 

For the specific heat and magnetic susceptibility, shown in Figs \ref{fig:cv} and \ref{fig:khi}, we see that
the peaks are around very close temperatures for all the lattice sizes so that a finite size scaling behavior
is not visible to the eyes. In fact the finite size scaling effects for the critical temperature are very 
subtle when one uses non multiple of three lattices in the BW model\cite{Jorge2016}. All these observations 
are therefore not conclusive about the order of the phase transition.

\subsection{Continuous phase transition}

In order to investigate if the system undergoes a continuous phase transition we carried
out a study seeking out for critical exponents and critical temperature. By locating the
maxima of the relations defined in Eqs. (\ref{eq:V1})-(\ref{eq:V6}), which scales as shown 
in Eq. (\ref{eq:Vj}), we find $1/\nu$ as the slopes of the straight lines of the plots of $V_j$ 
against $\ln L$, since in $T_c(t=0)$ the coefficients $\mathcal{V}_j$ are constants independent 
of the lattice size $L$.
\begin{figure} [h] 
\centering
\hspace{2.0in}
\includegraphics[scale=0.34,angle=-90]{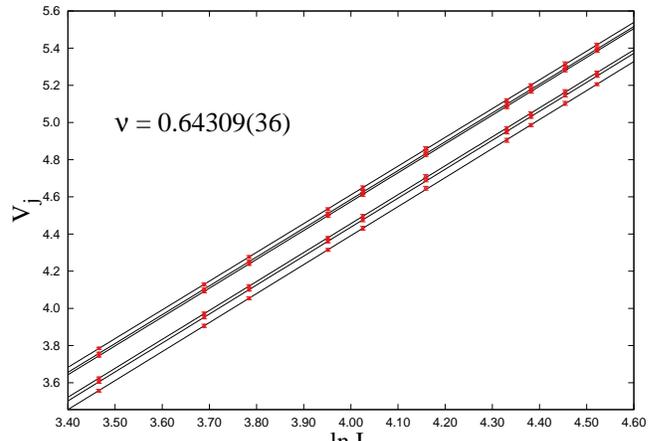}
\vspace{-0.20in}
\caption{Size dependence of $V_j$ at the critical temperature. The slopes yield $1/\nu$. \label{fig:ni}}
\end{figure}

Fig.\ref{fig:ni} shows the six slopes, where for each one of them was calculated $\nu=1/\left(\frac{1}{\nu}\right)$
with error $\varDelta \nu=\varDelta \left(\frac{1}{\nu}\right)/\left(\frac{1}{\nu}\right)^2$.
The graph presents only one of the ten sets of the performed finite size scaling simulations.
\begin{figure} [h] 
\centering
\hspace{2.0in}
%\vspace{0.1in}
%\includegraphics[scale=0.40]{afig1.eps}
\includegraphics[scale=0.34,angle=-90]{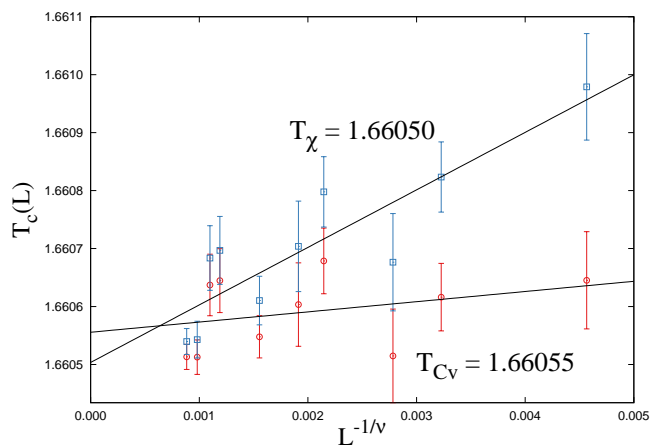}
\vspace{-0.20in}
\caption{Size dependence of the locations of the extrema in the specific heat and the magnetic 
susceptibility with $\nu=0.6438$. \label{fig:fit_tc}}
\end{figure}
\begin{figure} [t] 
\centering
\hspace{2.0in}
%\vspace{0.1in}
%\includegraphics[scale=0.40]{afig1.eps}
\includegraphics[scale=0.34,angle=-90]{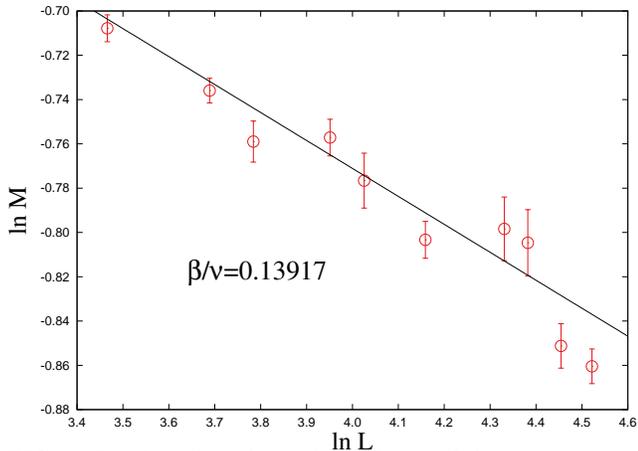}
\vspace{-0.20in}
\caption{Log-log plot of size dependence of the magnetization at $T_c=1.660549$. \label{fig:beta}}
\end{figure}
\begin{figure} [t] 
\centering
\hspace{2.0in}
%\vspace{0.1in}
%\includegraphics[scale=0.40]{afig1.eps}
\includegraphics[scale=0.34,angle=-90]{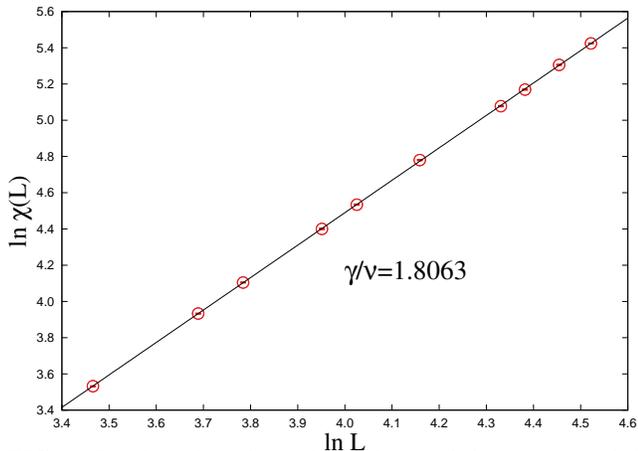}
\vspace{-0.20in}
\caption{Log-log plot of size dependence of the maxima of the susceptibility. \label{fig:gama}}
\end{figure}

The value obtained on each set is shown in the first column of Table \ref{table:resul} and the final
result $\nu=0.6438(10)$ appears in the last line.

With the exponent $\nu$ accurately determined we proceed to estimate the critical temperature and the
the exponents $\beta$ and $\gamma$. In Fig. \ref{fig:fit_tc} we use Eq. (\ref{eq:tc}) to determine $T_c$ 
as the extrapolation to $L\rightarrow\infty$ ($L^{-1/\nu}=0$) of the linear fits given by the locations 
of the maxima of the specific heat and the magnetic susceptibility. As expected the points look badly 
aligned because of the proximity of the points to the exact value. In Fig. \ref{fig:beta} we depict the
log-log plot of the magnetization at the critical temperature calculated by Eq. (\ref{eq:mag}). The slope
gives $\beta/\nu$, so that we obtain $\beta=\nu\frac{\beta}{\nu}$, with the error $\varDelta\beta=\frac{\beta}{\nu}
\varDelta\nu+\nu\varDelta\frac{\beta}{\nu}$, yielding $\beta=0.0896(69)$. 
Fig. \ref{fig:gama} displays the log-log plot of the maxima of the susceptibility, were we obtain $\gamma=1.1629(42)$.
In the last three columns of Table \ref{table:resul} we present the results for all the 10 sets for 
$\beta$, $\gamma$, and $T_c$, with the final mean values $\beta=0.0762(75)$, $\gamma=1.1611(28)$, and 
$T_c=1.660549(51)$ in the last line. These results with well defined critical exponents and critical 
temperature point out to a typical continuous phase transition corroborating the conclusions of Costa and Plascak\cite{Costa2004}. 
\begin{table}[h!]
   \centering
   \caption{Ten finite size scaling results for the critical 
   temperature, $T_c$, and the exponents $\nu$,   $\beta$ and
   $\gamma$. The averages over all runs are shown in the last 
   line. \label{table:resul}}
   
\begin{tabular} {c|c|c|c} 
\hline
  $\nu$ & $\beta$ & $\gamma$ & $T_c$  \\
\hline  
   0.64309(36) &  0.0896(69) & 1.1629(42) &  1.660469(26)\\
   0.64309(47) &  0.0854(82) & 1.1640(52) &  1.660480(34)\\
   0.64568(22) &  0.0760(56) & 1.1588(42) &  1.660556(25)\\
   0.64404(33) &  0.0722(74) & 1.1576(48) &  1.660568(32)\\
   0.64365(32) &  0.0663(35) & 1.1622(42) &  1.660607(15)\\
   0.64463(42) &  0.0667(54) & 1.1594(56) &  1.660624(21)\\
   0.64442(39) &  0.0734(58) & 1.1584(45) &  1.660575(24)\\
   0.64373(35) &  0.0761(54) & 1.1610(45) &  1.660556(23)\\
   0.64413(32) &  0.0815(78) & 1.1604(42) &  1.660504(29)\\
   0.64178(37) &  0.0753(78) & 1.1663(44) &  1.660552(33)\\
\hline                                      
   0.6438(10) &  0.0762(75) & 1.1611(28) &  1.660549(51)\\
\hline
\hline
\end{tabular}
\end{table}
\begin{figure} [h!] 
\centering
\hspace{2.0in}
%\vspace{0.1in}
%\includegraphics[scale=0.40]{afig1.eps}
\includegraphics[scale=0.34,angle=-90]{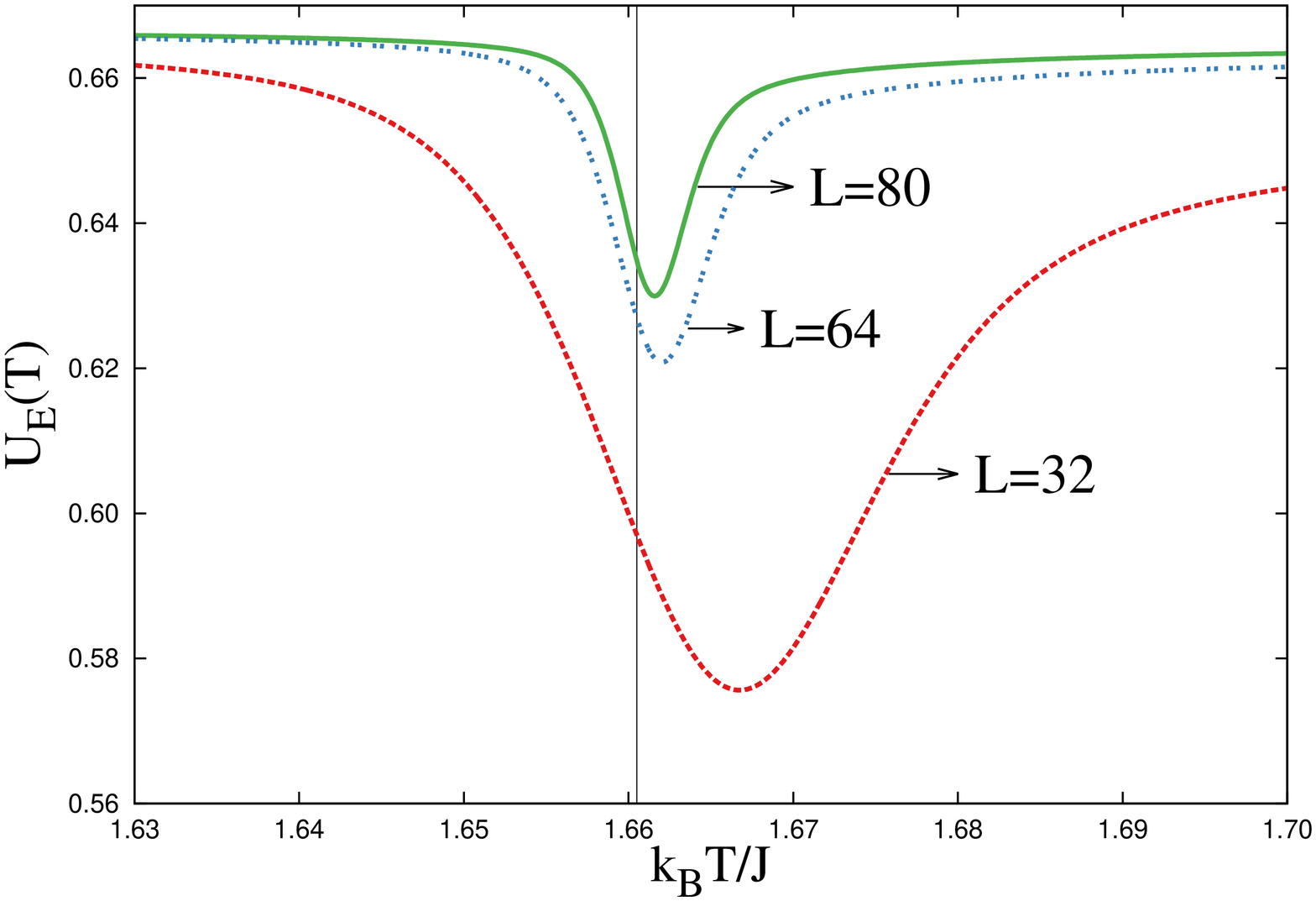}
\vspace{-0.20in}
\caption{Behavior of the cumulant of energy. The vertical line demarcates the critical temperature estimated using scaling laws 
for the continuous transition \label{fig:u4}}
\end{figure}
\begin{figure} [h!] 
\centering
\hspace{2.0in}
%\vspace{0.1in}
%\includegraphics[scale=0.40]{afig1.eps}
\includegraphics[scale=0.34,angle=-90]{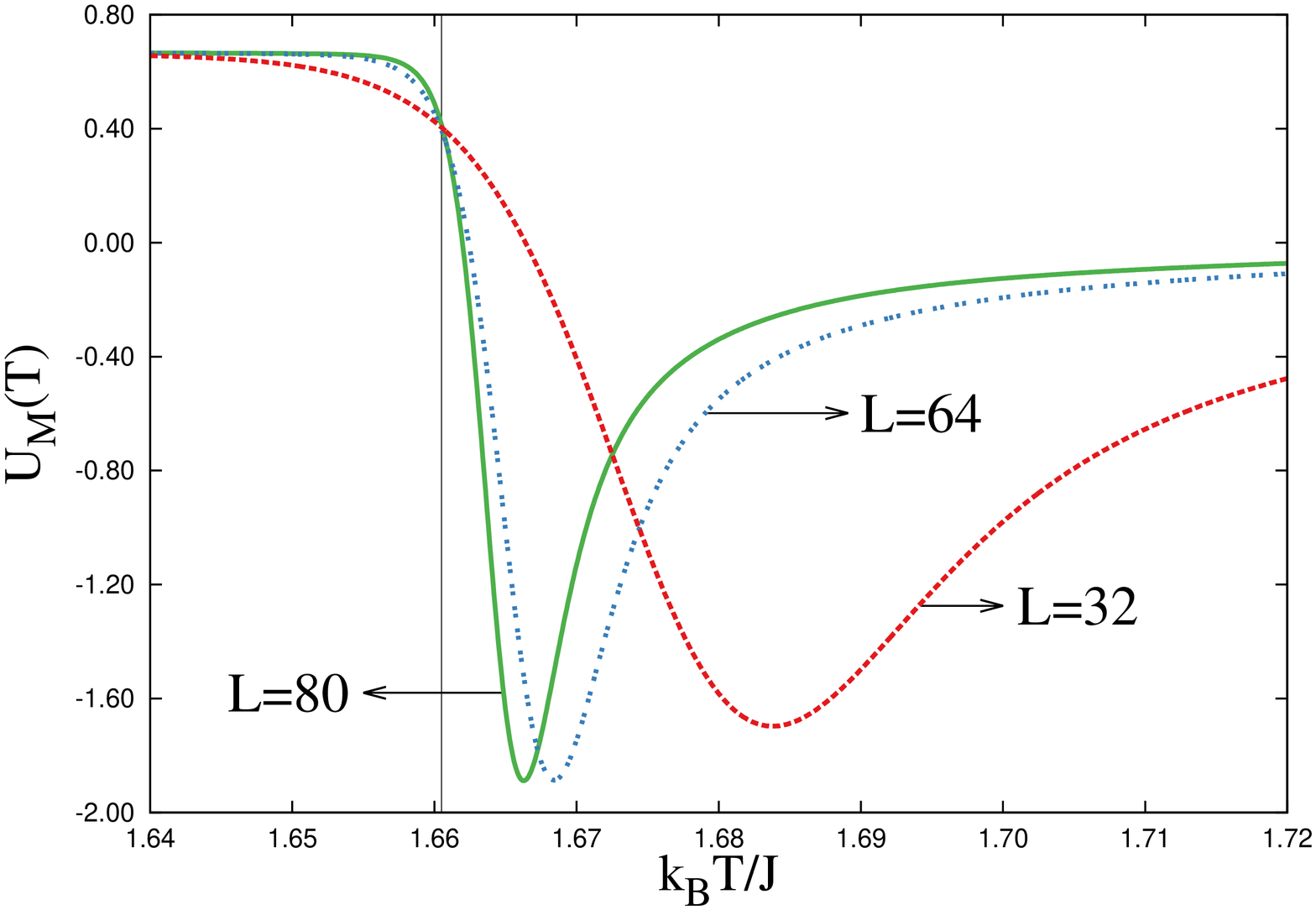}
\vspace{-0.20in}
\caption{Behavior of the cumulant of magnetization. The vertical line demarcates the critical temperature estimated using scaling laws 
for the continuous transition \label{fig:um4}}
\end{figure}
\begin{figure} [h!] 
\centering
\hspace{2.0in}
\vspace{0.1in}
\includegraphics[scale=0.34,angle=-90]{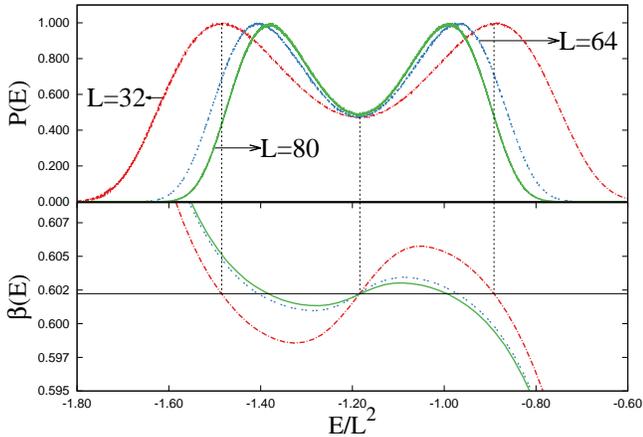}
\vspace{-0.20in}
\caption{(Top) Canonical distribution of $P(E)$. (Bottom) Microcanonical inverse temperature. The horizontal
line demarcates the critical temperature estimated using scaling laws for the continuous transition.
\label{fig:pe}}
\end{figure}
\begin{figure} [h!] 
\centering
\hspace{2.0in}
\vspace{0.1in}
\includegraphics[scale=0.34,angle=-90]{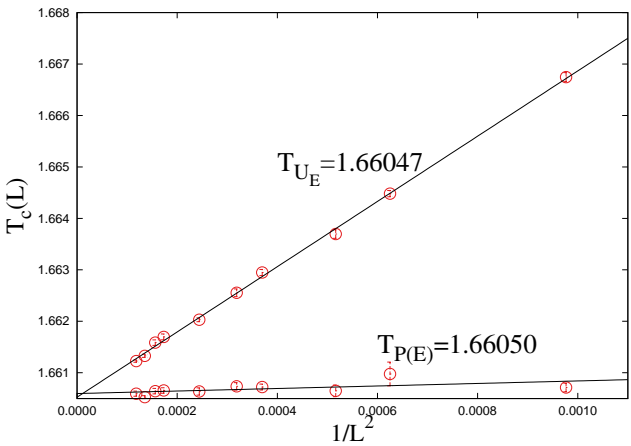}
\vspace{-0.20in}
\caption{Fitting of the temperatures of the minima of the cumulant of energy and of the temperatures where the 
same height in $P(E)$ occurs, determining the critical temperature.\label{fig:fit-tc-u4e} }
\end{figure}
\begin{figure} [h!] 
\centering
\hspace{2.0in}
\vspace{0.1in}
\includegraphics[scale=0.34,angle=-90]{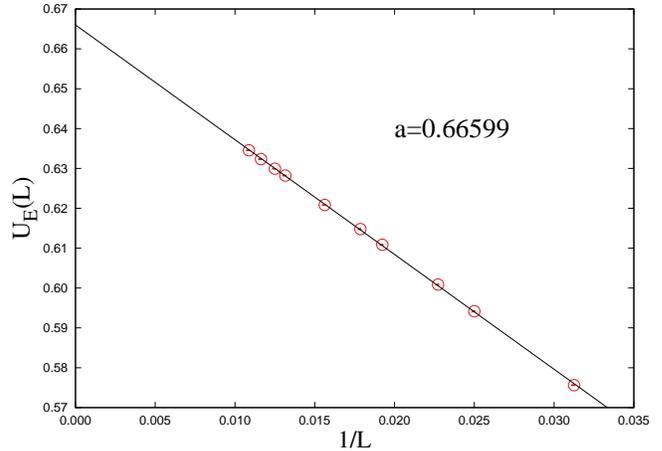}
\vspace{-0.20in}
\caption{Linear fitting of the minimum of the cumulant of energy with $1/L$. \label{fig:fit-u4e}}
\end{figure}

\begin{figure} [h] 
\centering
\hspace{2.0in}
\vspace{0.1in}
\includegraphics[scale=0.34,angle=-90]{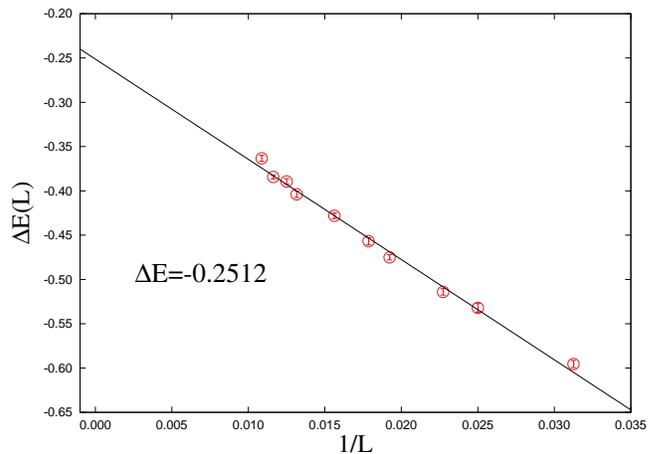}
\vspace{-0.20in}
\caption{Variation of the energy between the peaks of of the energy
probability against $1/L$. \label{fig:pe_de}}
\end{figure}

\begin{figure} [t!] 
\centering
\includegraphics[scale=0.3,angle=-90]{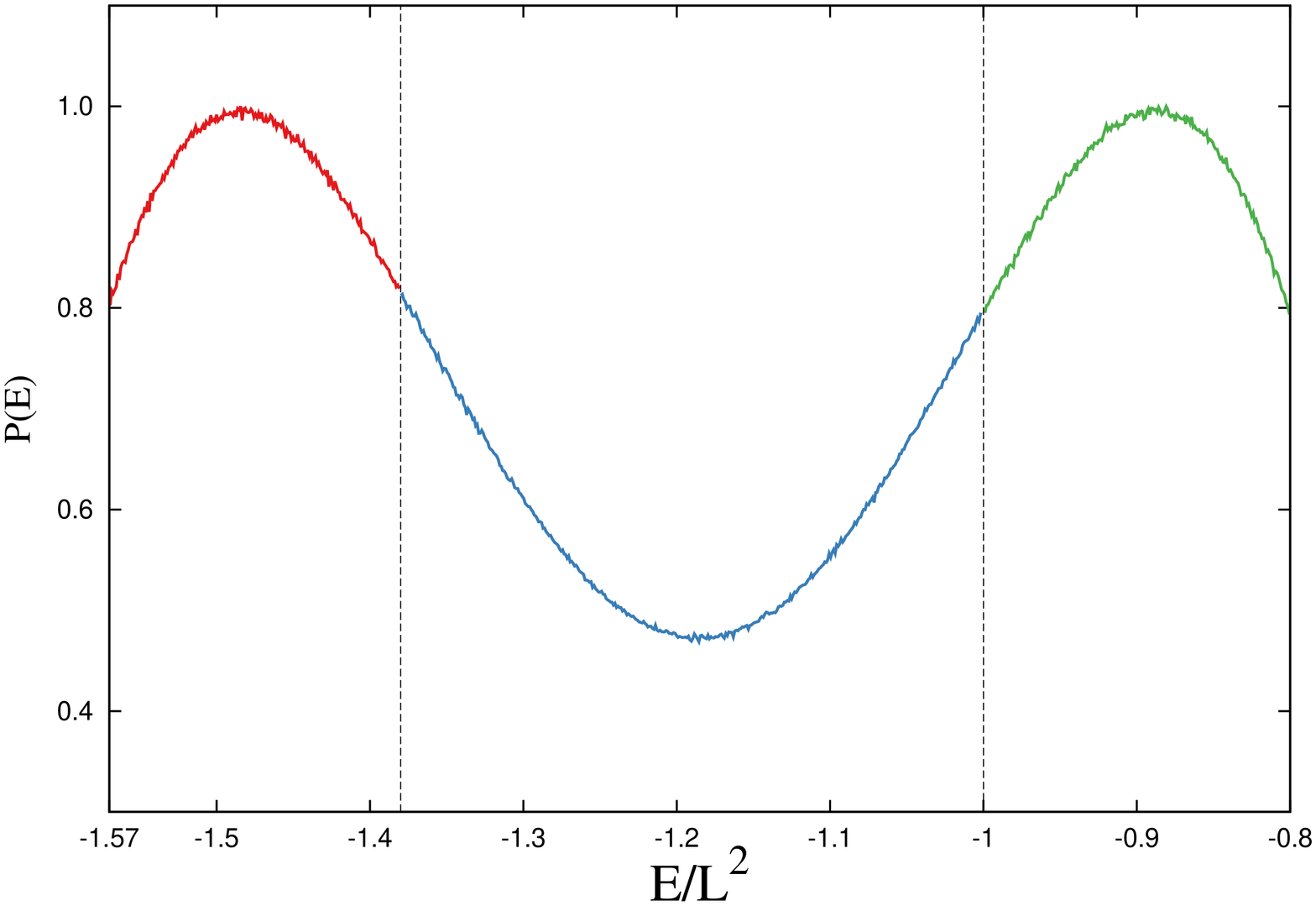}
\includegraphics[scale=0.36,angle=0]{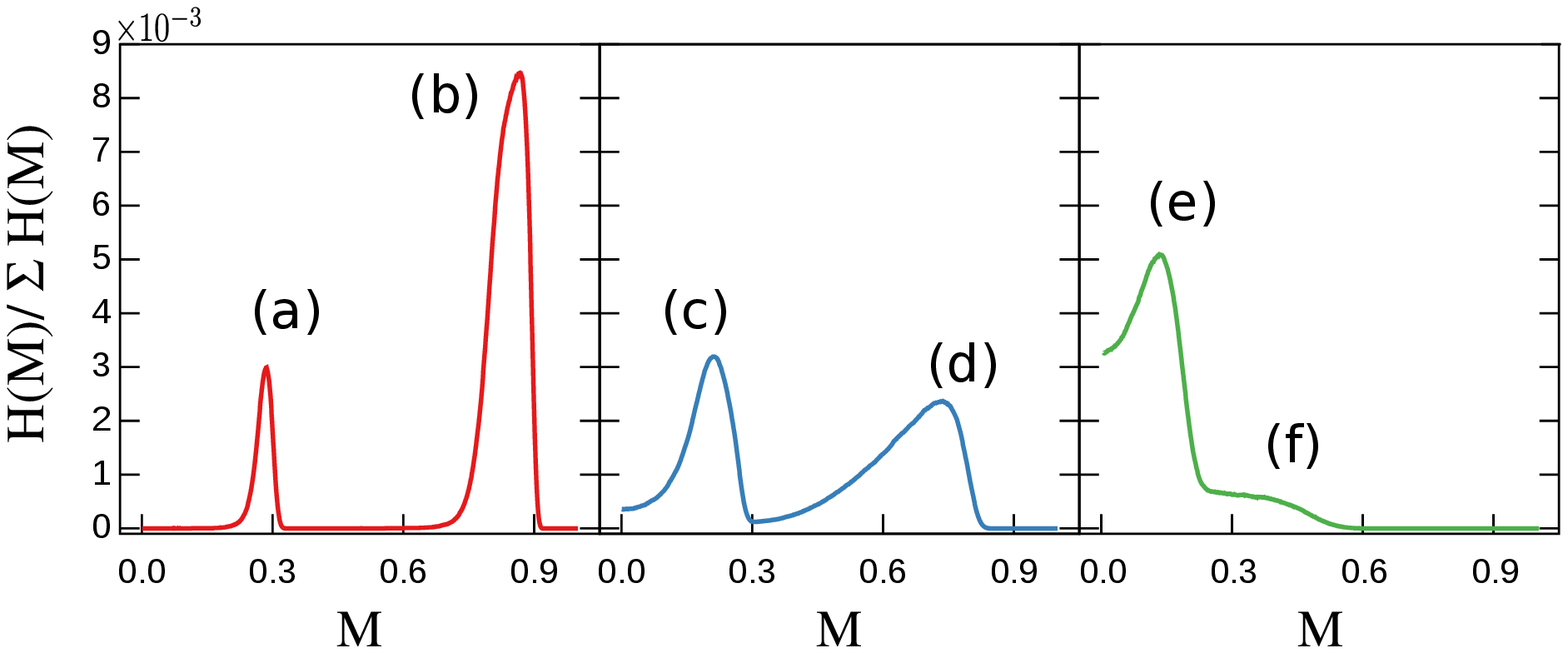}
\vspace{-0.3 cm}
\caption{(Top) Canonical Distribution of $P(E)$ at the temperature of same height. 
(Bottom) Histograms of the magnetization in the regions close to the peaks and the valley. \label{fig:config-r}}
\end{figure}

\begin{figure*} [t] 
\centering
\hspace{2.0in}
\vspace{0.1in}
\includegraphics[scale=0.70,angle=-90]{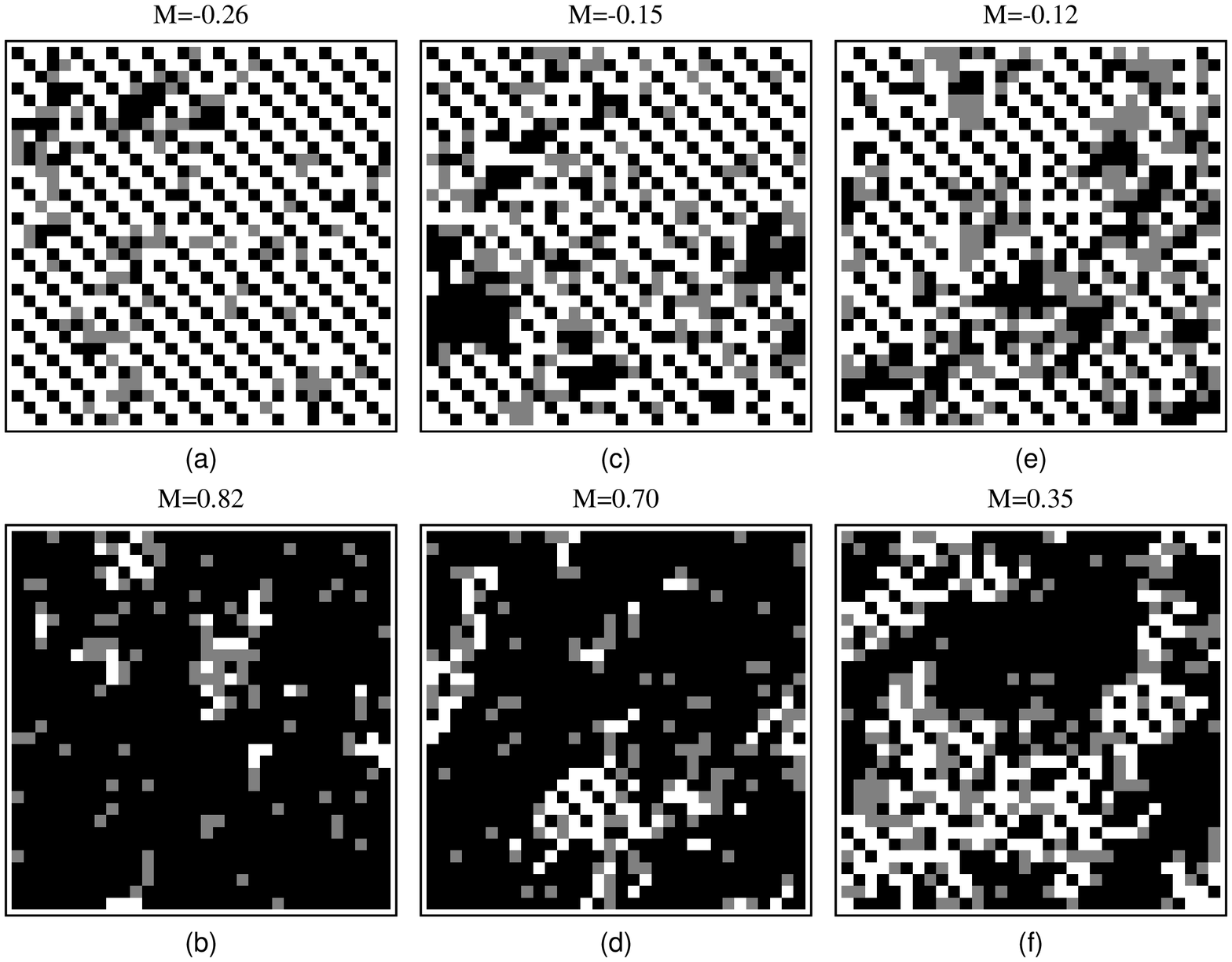}
\vspace{-0.20in}
\caption{Configurations corresponding to the peaks of magnetization in the histograms of Fig. \ref{fig:config-r} (bottom) . \label{fig:config-rgb}}
\end{figure*}

\subsection{Discontinuous phase transition}

Considering the system as undergoing a discontinuous phase transition we analyze the behavior
of the fourth-order Binder energy and magnetization cumulants, and the energy probability distribution.
In Fig. \ref{fig:u4} we show the energy cumulant for three lattice sizes $L=32,~64,$ e $80$.
One can see that the curves present a minimum in the region of the critical temperature and
the temperatures of these extrema approach the critical temperature, while the values of the
functions tend to something around $0.666$.

For the magnetization cumulant we have a typical discontinuous transition behavior characterized by a
prominent minimum and the crossing of the cumulants of different sizes at the critical temperature, as
shown in Fig. \ref{fig:um4} where we depict the cumulant of magnetization for $L=32,~64,$ e $80$. The vertical
line in Figs. \ref{fig:u4} and \ref{fig:um4} demarcates the critical temperature estimated using scaling laws 
for the continuous transition.

In Fig. \ref{fig:pe} we present two other quantities, which characterize a discontinuous transition
behavior: the energy probability and the inverse microcanonical temperature. In the top we show the
energy probability as a function of the energy per particle for three lattice sizes.

At the transition temperature we observe a double peak of same height and a valley between them with 
non-zero probability. In the bottom we display the microcanonical inverse temperature as a function 
of energy revealing a manifest discontinuous transition behavior. The
horizontal line represents the critical temperature obtained by means of scaling laws for a continuous
transition. One can see that the crossing of the curves occurs around the critical temperature and that
the minimum of the probability in the valley coincides with this point.

Using the scaling laws for the minimum of the energy cumulant and the energy probability Eq. \ref{eq:tc_first}, we can
estimate the critical temperature of the system. The linear fitting of these two quantities are represented in Fig. 
\ref{fig:fit-tc-u4e}. These are the results of the first of ten sets of samplings. One can see that the values agree
with those obtained by means of the analysis of a continuous transition using the specific heat and the susceptibility.

Two other additional results that corroborate the discontinuous phase transition are the linear
fitting of the minimum of the energy cumulant, which converges to $0.666...$ for an infinite
system, and the fitting of the variation of the energy between the peaks of of the energy probability,
the latent heat. We present these two quantities in Figs. \ref{fig:fit-u4e} and \ref{fig:pe_de}, 
respectively. In both situations one can see a typical discontinuous transition behavior.

In Fig. \ref{fig:fit-u4e} we show the fitting of the minimum of the energy cumulant
as a function of $1/L$ using $U_{E}(L,T_c)= a+bL^{-1}$. We see that 
$\lim _{ L\rightarrow \infty}U_{E}(L,T_c)=2/3$ as expected \cite{Lee1991}. In its turn
Fig. \ref{fig:pe_de} displays the fitting of the variation of energy between the 
peaks of the energy probability at the critical temperature as a function of $1/L$.
We see that as $L\rightarrow \infty$ the peaks keep an energy
difference of $|\Delta E| = 0.2512$, which reveals the existence
of a latent heat, which is characteristic of a discontinuous transition.

\subsection{Coexistence of continuous and discontinuous phase transitions}

So far we have analyzed the system adopting two different approaches: 
i) considering that it undergoes a continuous phase transition, and ii) the 
phase transition is discontinuous. In the first case we have obtained quantities
typical of this transition, such as the critical exponents $\nu=0.6438(10)$, 
$\beta= 0.0762(75)$ and $\gamma= 1.1611(28)$, and the critical temperature
$T_c=1.660549(51)$, as prescribed by \cite{Costa2004}, confirming therefore
that the system undergoes a continuous phase transition. It is also worthwhile to 
note that these results are very close to the exact solution for the BW spin $1/2$:
$\nu\cong0.6667$, $\beta\cong0.0833$, and $\gamma\cong1.1667$. On the other hand, 
assuming that the transition is discontinuous, the system behaves this way,
with a critical temperature $T_c= 1.66055(38)$, and a latent heat 
$|\Delta E| = 0.2512$, as described by Kinzel \cite{Kinzel1981}. We are thus 
faced with an indeterminacy, which is characteristic of a multicritical transition.
In such situation the question is: which are the phases present in the criticality?

In order to address this discussion we have to analyze the configurations that appear
at the criticality. First we have to define an interval of more probable energies at
the critical temperature. The energy probability at the critical temperature 
$P(E)=g(E)e^{-E/k_BT_c}$ displays two peaks revealing the two more probable energies, 
and a non-null valley between them. We can therefore define three energy intervals: 
one around the lower energy peak, a second between the peaks, and the last close to the 
highest energy peak, as shown in top of Fig. \ref{fig:config-r} for $L=32$. For each of these 
regions we constructed histograms of the magnetization, which are shown in 
bottom of Fig. \ref{fig:config-r}. 
In all cases we observe the presence of two peaks. The peak (a) is related to ferrimagnetic 
configurations with magnetization around $-L^2/3$, while the peak (b) correspond to 
ferromagnetic configurations with magnetization close to $L^2$, as shown in Fig. 
\ref{fig:config-rgb} (a) and (b). The black, white and gray squares represent the spins 
$+1$, $-1$ e $0$, respectively. We see that in Fig. \ref{fig:config-rgb}(a) we have the 
predominance of the ferrimagnetic configuration with some spins $0$ points and the occurrence 
of some paramagnetic blocks, and in Fig. \ref{fig:config-rgb}(b) the predominance of the 
ferromagnetic configuration, with some points of spins $0$ and the occurrence a few 
paramagnetic blocks. In its turn the peaks (c) and (d) of Fig. \ref{fig:config-r} correspond 
to configurations with the predominance of ferri- and ferromagnetic agglomerates, respectively, 
with the presence of spins $0$, some islands diverse of the predominant, and small paramagnetic 
blocks, as observed in Fig. \ref{fig:config-rgb} (c) and (d). Its worthwhile mentioning that in (d) 
there is a higher incidence of paramagnetic clusters then in (c). In peaks (e) and (f) we see 
that the system is closer to disorder, but there is still a noticeable presence of ferrimagnetic 
and ferromagnetic clusters in each of them, respectively.

It is important to notice that each ferrimagnetic cluster, as in Fig. \ref{fig:config-rgb} (a) can
belong to one of the three possible ground state ferrimagnetic configurations, in such a way that
all the situations described above correspond to the coexistence of three ferrimagnetic and a ferromagnetic 
configurations, along with the paramagnetic configuration indicating that we are dealing with a tetracritical 
order-disorder transition, since all these configurations coexist at the critical temperature. All these evidences, 
such as the indeterminacy of the order of the phase transition and the non-null valley in the energy probability,
point out to the behavior of a tetracritical point, which can correspond to the zero field state of
critical lines of external or crystalline fields.

All the discussion above led to a non-conclusive result on the order of the phase transition of the
spin-1 Baxter Wu model, suggesting a coexistence of the continuous and the discontinuous phase transitions. It
seams quite controversial, but a recent publication\cite{Post2018} shows the coexistence of first and 
second order electronic phase transition in a correlated oxide, where the sample bulk exhibits a first-order  
transition between metal and insulator phases, while the anomalous nanoscale domain walls in the insulating 
state undergo a distinctly continuous insulator-metal transition, with characteristics of second-order behavior.
Such example encourage us to suggest that the ferromagnetic state undergoes a discontinuous phase transition, 
while the ferrimagnetic states pass through a continuous phase transition.

\section{Conclusions}
We studied the spin-1 Baxter-Wu model either considering that it undergoes a continuous, either a discontinuous 
phase transition. In both cases the results have lead to strong evidences favoring the understanding that the 
transition is continuous or discontinuous. An analysis of the configurations in the region of the criticality
shows the coexistence of ferromagnetic, ferrimagnetic, and paramagnetic agglomerates in different configurations
belonging to the criticality, revealing a tetracritical behavior, since the model displays three alternative 
ferrimagnetic constructions. As a result the system may be considered as the zero field multicritical point of
possible critical lines of external or crystalline fields. 

\section{Acknowledgements}

We acknowledge the computer resources provided by LCC-UFG and IF-UFMT. L. N. 
Jorge and L. S. Ferreira acknowledge the support by FAPEG and CAPES, respectively.


\begin{thebibliography}{42}%
\makeatletter
\providecommand \@ifxundefined [1]{%
 \@ifx{#1\undefined}
}%
\providecommand \@ifnum [1]{%
 \ifnum #1\expandafter \@firstoftwo
 \else \expandafter \@secondoftwo
 \fi
}%
\providecommand \@ifx [1]{%
 \ifx #1\expandafter \@firstoftwo
 \else \expandafter \@secondoftwo
 \fi
}%
\providecommand \natexlab [1]{#1}%
\providecommand \enquote  [1]{``#1''}%
\providecommand \bibnamefont  [1]{#1}%
\providecommand \bibfnamefont [1]{#1}%
\providecommand \citenamefont [1]{#1}%
\providecommand \href@noop [0]{\@secondoftwo}%
\providecommand \href [0]{\begingroup \@sanitize@url \@href}%
\providecommand \@href[1]{\@@startlink{#1}\@@href}%
\providecommand \@@href[1]{\endgroup#1\@@endlink}%
\providecommand \@sanitize@url [0]{\catcode `\\12\catcode `\$12\catcode
  `\&12\catcode `\#12\catcode `\^12\catcode `\_12\catcode `\%12\relax}%
\providecommand \@@startlink[1]{}%
\providecommand \@@endlink[0]{}%
\providecommand \url  [0]{\begingroup\@sanitize@url \@url }%
\providecommand \@url [1]{\endgroup\@href {#1}{\urlprefix }}%
\providecommand \urlprefix  [0]{URL }%
\providecommand \Eprint [0]{\href }%
\providecommand \doibase [0]{http://dx.doi.org/}%
\providecommand \selectlanguage [0]{\@gobble}%
\providecommand \bibinfo  [0]{\@secondoftwo}%
\providecommand \bibfield  [0]{\@secondoftwo}%
\providecommand \translation [1]{[#1]}%
\providecommand \BibitemOpen [0]{}%
\providecommand \bibitemStop [0]{}%
\providecommand \bibitemNoStop [0]{.\EOS\space}%
\providecommand \EOS [0]{\spacefactor3000\relax}%
\providecommand \BibitemShut  [1]{\csname bibitem#1\endcsname}%
\let\auto@bib@innerbib\@empty
%</preamble>
\bibitem [{\citenamefont {Roelofs}\ \emph {et~al.}(1981)\citenamefont
  {Roelofs}, \citenamefont {Kortan}, \citenamefont {Einstein},\ and\
  \citenamefont {Park}}]{Roelofs1981}%
  \BibitemOpen
  \bibfield  {author} {\bibinfo {author} {\bibfnamefont {L.~D.}\ \bibnamefont
  {Roelofs}}, \bibinfo {author} {\bibfnamefont {A.}~\bibnamefont {Kortan}},
  \bibinfo {author} {\bibfnamefont {T.}~\bibnamefont {Einstein}}, \ and\
  \bibinfo {author} {\bibfnamefont {R.~L.}\ \bibnamefont {Park}},\ }\href@noop
  {} {\bibfield  {journal} {\bibinfo  {journal} {Physical Review Letters}\
  }\textbf {\bibinfo {volume} {46}},\ \bibinfo {pages} {1465} (\bibinfo {year}
  {1981})}\BibitemShut {NoStop}%
\bibitem [{\citenamefont {Piercy}\ and\ \citenamefont
  {Pfn{\"u}r}(1987)}]{Piercy1987}%
  \BibitemOpen
  \bibfield  {author} {\bibinfo {author} {\bibfnamefont {P.}~\bibnamefont
  {Piercy}}\ and\ \bibinfo {author} {\bibfnamefont {H.}~\bibnamefont
  {Pfn{\"u}r}},\ }\href@noop {} {\bibfield  {journal} {\bibinfo  {journal}
  {Physical review letters}\ }\textbf {\bibinfo {volume} {59}},\ \bibinfo
  {pages} {1124} (\bibinfo {year} {1987})}\BibitemShut {NoStop}%
\bibitem [{\citenamefont {Schwenger}\ \emph {et~al.}(1994)\citenamefont
  {Schwenger}, \citenamefont {Budde}, \citenamefont {Voges},\ and\
  \citenamefont {Pfn{\"u}r}}]{Schwenger1994}%
  \BibitemOpen
  \bibfield  {author} {\bibinfo {author} {\bibfnamefont {L.}~\bibnamefont
  {Schwenger}}, \bibinfo {author} {\bibfnamefont {K.}~\bibnamefont {Budde}},
  \bibinfo {author} {\bibfnamefont {C.}~\bibnamefont {Voges}}, \ and\ \bibinfo
  {author} {\bibfnamefont {H.}~\bibnamefont {Pfn{\"u}r}},\ }\href@noop {}
  {\bibfield  {journal} {\bibinfo  {journal} {Physical review letters}\
  }\textbf {\bibinfo {volume} {73}},\ \bibinfo {pages} {296} (\bibinfo {year}
  {1994})}\BibitemShut {NoStop}%
\bibitem [{\citenamefont {Wood}\ and\ \citenamefont
  {Griffiths}(1972)}]{Wood1972}%
  \BibitemOpen
  \bibfield  {author} {\bibinfo {author} {\bibfnamefont {D.}~\bibnamefont
  {Wood}}\ and\ \bibinfo {author} {\bibfnamefont {H.}~\bibnamefont
  {Griffiths}},\ }\href@noop {} {\bibfield  {journal} {\bibinfo  {journal}
  {Journal of Physics C: Solid State Physics}\ }\textbf {\bibinfo {volume}
  {5}},\ \bibinfo {pages} {L253} (\bibinfo {year} {1972})}\BibitemShut
  {NoStop}%
\bibitem [{\citenamefont {Baxter}\ and\ \citenamefont {Wu}(1973)}]{Baxter1973}%
  \BibitemOpen
  \bibfield  {author} {\bibinfo {author} {\bibfnamefont {R.~J.}\ \bibnamefont
  {Baxter}}\ and\ \bibinfo {author} {\bibfnamefont {F.}~\bibnamefont {Wu}},\
  }\href@noop {} {\bibfield  {journal} {\bibinfo  {journal} {Physical Review
  Letters}\ }\textbf {\bibinfo {volume} {31}},\ \bibinfo {pages} {1294}
  (\bibinfo {year} {1973})}\BibitemShut {NoStop}%
\bibitem [{\citenamefont {Baxter}(1974)}]{Baxter1974}%
  \BibitemOpen
  \bibfield  {author} {\bibinfo {author} {\bibfnamefont {R.}~\bibnamefont
  {Baxter}},\ }\href@noop {} {\bibfield  {journal} {\bibinfo  {journal}
  {Australian Journal of Physics}\ }\textbf {\bibinfo {volume} {27}},\ \bibinfo
  {pages} {369} (\bibinfo {year} {1974})}\BibitemShut {NoStop}%
\bibitem [{\citenamefont {Baxter}\ and\ \citenamefont
  {Wu}(1974)}]{Baxter1974a}%
  \BibitemOpen
  \bibfield  {author} {\bibinfo {author} {\bibfnamefont {R.}~\bibnamefont
  {Baxter}}\ and\ \bibinfo {author} {\bibfnamefont {F.}~\bibnamefont {Wu}},\
  }\href@noop {} {\bibfield  {journal} {\bibinfo  {journal} {Australian Journal
  of Physics}\ }\textbf {\bibinfo {volume} {27}},\ \bibinfo {pages} {357}
  (\bibinfo {year} {1974})}\BibitemShut {NoStop}%
\bibitem [{\citenamefont {Santos}\ and\ \citenamefont
  {Figueiredo}(2001)}]{Santos2001}%
  \BibitemOpen
  \bibfield  {author} {\bibinfo {author} {\bibfnamefont {M.}~\bibnamefont
  {Santos}}\ and\ \bibinfo {author} {\bibfnamefont {W.}~\bibnamefont
  {Figueiredo}},\ }\href@noop {} {\bibfield  {journal} {\bibinfo  {journal}
  {Physical Review E}\ }\textbf {\bibinfo {volume} {63}},\ \bibinfo {pages}
  {042101} (\bibinfo {year} {2001})}\BibitemShut {NoStop}%
\bibitem [{\citenamefont {Caparica}\ \emph {et~al.}(2015)\citenamefont
  {Caparica}, \citenamefont {Leão},\ and\ \citenamefont
  {DaSilva}}]{Caparica2015b}%
  \BibitemOpen
  \bibfield  {author} {\bibinfo {author} {\bibfnamefont {A.}~\bibnamefont
  {Caparica}}, \bibinfo {author} {\bibfnamefont {S.~A.}\ \bibnamefont {Leão}},
  \ and\ \bibinfo {author} {\bibfnamefont {C.~J.}\ \bibnamefont {DaSilva}},\
  }\href {\doibase http://dx.doi.org/10.1016/j.physa.2015.06.002} {\bibfield
  {journal} {\bibinfo  {journal} {Physica A: Statistical Mechanics and its
  Applications}\ }\textbf {\bibinfo {volume} {438}},\ \bibinfo {pages} {447 }
  (\bibinfo {year} {2015})}\BibitemShut {NoStop}%
\bibitem [{\citenamefont {Kinzel}\ \emph {et~al.}(1981)\citenamefont {Kinzel},
  \citenamefont {Domany},\ and\ \citenamefont {Aharony}}]{Kinzel1981}%
  \BibitemOpen
  \bibfield  {author} {\bibinfo {author} {\bibfnamefont {W.}~\bibnamefont
  {Kinzel}}, \bibinfo {author} {\bibfnamefont {E.}~\bibnamefont {Domany}}, \
  and\ \bibinfo {author} {\bibfnamefont {A.}~\bibnamefont {Aharony}},\ }\href
  {http://stacks.iop.org/0305-4470/14/i=10/a=007} {\bibfield  {journal}
  {\bibinfo  {journal} {Journal of Physics A: Mathematical and General}\
  }\textbf {\bibinfo {volume} {14}},\ \bibinfo {pages} {L417} (\bibinfo {year}
  {1981})}\BibitemShut {NoStop}%
\bibitem [{\citenamefont {Costa}\ \emph {et~al.}(2004)\citenamefont {Costa},
  \citenamefont {Xavier},\ and\ \citenamefont {Plascak}}]{Costa2004}%
  \BibitemOpen
  \bibfield  {author} {\bibinfo {author} {\bibfnamefont {M.~L.~M.}\
  \bibnamefont {Costa}}, \bibinfo {author} {\bibfnamefont {J.~C.}\ \bibnamefont
  {Xavier}}, \ and\ \bibinfo {author} {\bibfnamefont {J.~A.}\ \bibnamefont
  {Plascak}},\ }\href {\doibase 10.1103/PhysRevB.69.104103} {\bibfield
  {journal} {\bibinfo  {journal} {Phys. Rev. B}\ }\textbf {\bibinfo {volume}
  {69}},\ \bibinfo {pages} {104103} (\bibinfo {year} {2004})}\BibitemShut
  {NoStop}%
\bibitem [{\citenamefont {Riedel}(1972)}]{Riedel1972}%
  \BibitemOpen
  \bibfield  {author} {\bibinfo {author} {\bibfnamefont {E.~K.}\ \bibnamefont
  {Riedel}},\ }\href {\doibase 10.1103/PhysRevLett.28.675} {\bibfield
  {journal} {\bibinfo  {journal} {Phys. Rev. Lett.}\ }\textbf {\bibinfo
  {volume} {28}},\ \bibinfo {pages} {675} (\bibinfo {year} {1972})}\BibitemShut
  {NoStop}%
\bibitem [{\citenamefont {Griffiths}(1973)}]{Griffiths1973}%
  \BibitemOpen
  \bibfield  {author} {\bibinfo {author} {\bibfnamefont {R.~B.}\ \bibnamefont
  {Griffiths}},\ }\href {\doibase 10.1103/PhysRevB.7.545} {\bibfield  {journal}
  {\bibinfo  {journal} {Phys. Rev. B}\ }\textbf {\bibinfo {volume} {7}},\
  \bibinfo {pages} {545} (\bibinfo {year} {1973})}\BibitemShut {NoStop}%
\bibitem [{\citenamefont {Graf}\ \emph {et~al.}(1967)\citenamefont {Graf},
  \citenamefont {Lee},\ and\ \citenamefont {Reppy}}]{Graf1967}%
  \BibitemOpen
  \bibfield  {author} {\bibinfo {author} {\bibfnamefont {E.~H.}\ \bibnamefont
  {Graf}}, \bibinfo {author} {\bibfnamefont {D.~M.}\ \bibnamefont {Lee}}, \
  and\ \bibinfo {author} {\bibfnamefont {J.~D.}\ \bibnamefont {Reppy}},\ }\href
  {\doibase 10.1103/PhysRevLett.19.417} {\bibfield  {journal} {\bibinfo
  {journal} {Phys. Rev. Lett.}\ }\textbf {\bibinfo {volume} {19}},\ \bibinfo
  {pages} {417} (\bibinfo {year} {1967})}\BibitemShut {NoStop}%
\bibitem [{\citenamefont {Landau}\ \emph {et~al.}(1971)\citenamefont {Landau},
  \citenamefont {Keen}, \citenamefont {Schneider},\ and\ \citenamefont
  {Wolf}}]{Landau1971}%
  \BibitemOpen
  \bibfield  {author} {\bibinfo {author} {\bibfnamefont {D.~P.}\ \bibnamefont
  {Landau}}, \bibinfo {author} {\bibfnamefont {B.~E.}\ \bibnamefont {Keen}},
  \bibinfo {author} {\bibfnamefont {B.}~\bibnamefont {Schneider}}, \ and\
  \bibinfo {author} {\bibfnamefont {W.~P.}\ \bibnamefont {Wolf}},\ }\href
  {\doibase 10.1103/PhysRevB.3.2310} {\bibfield  {journal} {\bibinfo  {journal}
  {Phys. Rev. B}\ }\textbf {\bibinfo {volume} {3}},\ \bibinfo {pages} {2310}
  (\bibinfo {year} {1971})}\BibitemShut {NoStop}%
\bibitem [{\citenamefont {Garland}\ and\ \citenamefont
  {Weiner}(1971)}]{Garland1971}%
  \BibitemOpen
  \bibfield  {author} {\bibinfo {author} {\bibfnamefont {C.~W.}\ \bibnamefont
  {Garland}}\ and\ \bibinfo {author} {\bibfnamefont {B.~B.}\ \bibnamefont
  {Weiner}},\ }\href {\doibase 10.1103/PhysRevB.3.1634} {\bibfield  {journal}
  {\bibinfo  {journal} {Phys. Rev. B}\ }\textbf {\bibinfo {volume} {3}},\
  \bibinfo {pages} {1634} (\bibinfo {year} {1971})}\BibitemShut {NoStop}%
\bibitem [{\citenamefont {Saul}\ \emph {et~al.}(1974)\citenamefont {Saul},
  \citenamefont {Wortis},\ and\ \citenamefont {Stauffer}}]{Saul1974}%
  \BibitemOpen
  \bibfield  {author} {\bibinfo {author} {\bibfnamefont {D.~M.}\ \bibnamefont
  {Saul}}, \bibinfo {author} {\bibfnamefont {M.}~\bibnamefont {Wortis}}, \ and\
  \bibinfo {author} {\bibfnamefont {D.}~\bibnamefont {Stauffer}},\ }\href
  {\doibase 10.1103/PhysRevB.9.4964} {\bibfield  {journal} {\bibinfo  {journal}
  {Phys. Rev. B}\ }\textbf {\bibinfo {volume} {9}},\ \bibinfo {pages} {4964}
  (\bibinfo {year} {1974})}\BibitemShut {NoStop}%
\bibitem [{\citenamefont {Musia\l{}}(2004)}]{Musial2004}%
  \BibitemOpen
  \bibfield  {author} {\bibinfo {author} {\bibfnamefont {G.}~\bibnamefont
  {Musia\l{}}},\ }\href {\doibase 10.1103/PhysRevB.69.024407} {\bibfield
  {journal} {\bibinfo  {journal} {Phys. Rev. B}\ }\textbf {\bibinfo {volume}
  {69}},\ \bibinfo {pages} {024407} (\bibinfo {year} {2004})}\BibitemShut
  {NoStop}%
\bibitem [{\citenamefont {{\v{S}}im{\.e}nas}\ \emph {et~al.}(2014)\citenamefont
  {{\v{S}}im{\.e}nas}, \citenamefont {Ibenskas},\ and\ \citenamefont
  {Tornau}}]{Simenas2014}%
  \BibitemOpen
  \bibfield  {author} {\bibinfo {author} {\bibfnamefont {M.}~\bibnamefont
  {{\v{S}}im{\.e}nas}}, \bibinfo {author} {\bibfnamefont {A.}~\bibnamefont
  {Ibenskas}}, \ and\ \bibinfo {author} {\bibfnamefont {E.~E.}\ \bibnamefont
  {Tornau}},\ }\href@noop {} {\bibfield  {journal} {\bibinfo  {journal}
  {Physical Review E}\ }\textbf {\bibinfo {volume} {90}},\ \bibinfo {pages}
  {042124} (\bibinfo {year} {2014})}\BibitemShut {NoStop}%
\bibitem [{\citenamefont {Fytas}(2011)}]{Fytas2011a}%
  \BibitemOpen
  \bibfield  {author} {\bibinfo {author} {\bibfnamefont {N.~G.}\ \bibnamefont
  {Fytas}},\ }\href {\doibase 10.1140/epjb/e2010-10738-y} {\bibfield  {journal}
  {\bibinfo  {journal} {The European Physical Journal B}\ }\textbf {\bibinfo
  {volume} {79}},\ \bibinfo {pages} {21} (\bibinfo {year} {2011})}\BibitemShut
  {NoStop}%
\bibitem [{\citenamefont {Dias}\ \emph {et~al.}(2017)\citenamefont {Dias},
  \citenamefont {Xavier},\ and\ \citenamefont {Plascak}}]{Dias2017}%
  \BibitemOpen
  \bibfield  {author} {\bibinfo {author} {\bibfnamefont {D.~A.}\ \bibnamefont
  {Dias}}, \bibinfo {author} {\bibfnamefont {J.~C.}\ \bibnamefont {Xavier}}, \
  and\ \bibinfo {author} {\bibfnamefont {J.~A.}\ \bibnamefont {Plascak}},\
  }\href {\doibase 10.1103/PhysRevE.95.012103} {\bibfield  {journal} {\bibinfo
  {journal} {Phys. Rev. E}\ }\textbf {\bibinfo {volume} {95}},\ \bibinfo
  {pages} {012103} (\bibinfo {year} {2017})}\BibitemShut {NoStop}%
\bibitem [{\citenamefont {Jorge}\ \emph {et~al.}(2016)\citenamefont {Jorge},
  \citenamefont {Ferreira}, \citenamefont {Le{\~a}o},\ and\ \citenamefont
  {Caparica}}]{Jorge2016}%
  \BibitemOpen
  \bibfield  {author} {\bibinfo {author} {\bibfnamefont {L.~N.}\ \bibnamefont
  {Jorge}}, \bibinfo {author} {\bibfnamefont {L.~S.}\ \bibnamefont {Ferreira}},
  \bibinfo {author} {\bibfnamefont {S.~A.}\ \bibnamefont {Le{\~a}o}}, \ and\
  \bibinfo {author} {\bibfnamefont {A.~A.}\ \bibnamefont {Caparica}},\ }\href
  {\doibase 10.1007/s13538-016-0439-y} {\bibfield  {journal} {\bibinfo
  {journal} {Brazilian Journal of Physics}\ }\textbf {\bibinfo {volume} {46}},\
  \bibinfo {pages} {556} (\bibinfo {year} {2016})}\BibitemShut {NoStop}%
\bibitem [{\citenamefont {Fisher}(1971)}]{Fisher1971}%
  \BibitemOpen
  \bibfield  {author} {\bibinfo {author} {\bibfnamefont {M.~E.}\ \bibnamefont
  {Fisher}},\ }\href@noop {} {\emph {\bibinfo {title} {Critical Phenomena}}},\
  edited by\ \bibinfo {editor} {\bibfnamefont {M.~S.}\ \bibnamefont {Green}}\
  (\bibinfo  {publisher} {Academic Press},\ \bibinfo {year} {1971})\BibitemShut
  {NoStop}%
\bibitem [{\citenamefont {Fisher}\ and\ \citenamefont
  {Barber}(1972)}]{Fisher1972}%
  \BibitemOpen
  \bibfield  {author} {\bibinfo {author} {\bibfnamefont {M.~E.}\ \bibnamefont
  {Fisher}}\ and\ \bibinfo {author} {\bibfnamefont {M.~N.}\ \bibnamefont
  {Barber}},\ }\href {\doibase 10.1103/PhysRevLett.28.1516} {\bibfield
  {journal} {\bibinfo  {journal} {Phys. Rev. Lett.}\ }\textbf {\bibinfo
  {volume} {28}},\ \bibinfo {pages} {1516} (\bibinfo {year}
  {1972})}\BibitemShut {NoStop}%
\bibitem [{\citenamefont {Ferrenberg}\ and\ \citenamefont
  {Landau}(1991)}]{Ferrenberg1991}%
  \BibitemOpen
  \bibfield  {author} {\bibinfo {author} {\bibfnamefont {A.~M.}\ \bibnamefont
  {Ferrenberg}}\ and\ \bibinfo {author} {\bibfnamefont {D.~P.}\ \bibnamefont
  {Landau}},\ }\href {\doibase 10.1103/PhysRevB.44.5081} {\bibfield  {journal}
  {\bibinfo  {journal} {Phys. Rev. B}\ }\textbf {\bibinfo {volume} {44}},\
  \bibinfo {pages} {5081} (\bibinfo {year} {1991})}\BibitemShut {NoStop}%
\bibitem [{\citenamefont {Caparica}\ \emph {et~al.}(2000)\citenamefont
  {Caparica}, \citenamefont {Bunker},\ and\ \citenamefont
  {Landau}}]{Caparica2000}%
  \BibitemOpen
  \bibfield  {author} {\bibinfo {author} {\bibfnamefont {A.}~\bibnamefont
  {Caparica}}, \bibinfo {author} {\bibfnamefont {A.}~\bibnamefont {Bunker}}, \
  and\ \bibinfo {author} {\bibfnamefont {D.}~\bibnamefont {Landau}},\
  }\href@noop {} {\bibfield  {journal} {\bibinfo  {journal} {Physical Review
  B}\ }\textbf {\bibinfo {volume} {62}},\ \bibinfo {pages} {9458} (\bibinfo
  {year} {2000})}\BibitemShut {NoStop}%
\bibitem [{\citenamefont {Challa}\ \emph {et~al.}(1986)\citenamefont {Challa},
  \citenamefont {Landau},\ and\ \citenamefont {Binder}}]{Challa1986}%
  \BibitemOpen
  \bibfield  {author} {\bibinfo {author} {\bibfnamefont {M.~S.~S.}\
  \bibnamefont {Challa}}, \bibinfo {author} {\bibfnamefont {D.~P.}\
  \bibnamefont {Landau}}, \ and\ \bibinfo {author} {\bibfnamefont
  {K.}~\bibnamefont {Binder}},\ }\href {\doibase 10.1103/PhysRevB.34.1841}
  {\bibfield  {journal} {\bibinfo  {journal} {Phys. Rev. B}\ }\textbf {\bibinfo
  {volume} {34}},\ \bibinfo {pages} {1841} (\bibinfo {year}
  {1986})}\BibitemShut {NoStop}%
\bibitem [{\citenamefont {Binder}\ and\ \citenamefont
  {Heermann}(2010)}]{Binder2010a}%
  \BibitemOpen
  \bibfield  {author} {\bibinfo {author} {\bibfnamefont {K.}~\bibnamefont
  {Binder}}\ and\ \bibinfo {author} {\bibfnamefont {D.}~\bibnamefont
  {Heermann}},\ }\href@noop {} {\emph {\bibinfo {title} {{M}onte {C}arlo
  simulation in statistical physics: an introduction}}}\ (\bibinfo  {publisher}
  {Springer Science \& Business Media},\ \bibinfo {year} {2010})\BibitemShut
  {NoStop}%
\bibitem [{\citenamefont {Vollmayr}\ \emph {et~al.}(1993)\citenamefont
  {Vollmayr}, \citenamefont {Reger}, \citenamefont {Scheucher},\ and\
  \citenamefont {Binder}}]{Vollmayr1993}%
  \BibitemOpen
  \bibfield  {author} {\bibinfo {author} {\bibfnamefont {K.}~\bibnamefont
  {Vollmayr}}, \bibinfo {author} {\bibfnamefont {J.~D.}\ \bibnamefont {Reger}},
  \bibinfo {author} {\bibfnamefont {M.}~\bibnamefont {Scheucher}}, \ and\
  \bibinfo {author} {\bibfnamefont {K.}~\bibnamefont {Binder}},\ }\href@noop {}
  {\bibfield  {journal} {\bibinfo  {journal} {Zeitschrift f{\"u}r Physik B
  Condensed Matter}\ }\textbf {\bibinfo {volume} {91}},\ \bibinfo {pages} {113}
  (\bibinfo {year} {1993})}\BibitemShut {NoStop}%
\bibitem [{\citenamefont {Binder}(1997)}]{Binder1997}%
  \BibitemOpen
  \bibfield  {author} {\bibinfo {author} {\bibfnamefont {K.}~\bibnamefont
  {Binder}},\ }\href@noop {} {\bibfield  {journal} {\bibinfo  {journal}
  {Reports on Progress in Physics}\ }\textbf {\bibinfo {volume} {60}},\
  \bibinfo {pages} {487} (\bibinfo {year} {1997})}\BibitemShut {NoStop}%
\bibitem [{\citenamefont {Lee}\ and\ \citenamefont
  {Kosterlitz}(1991)}]{Lee1991}%
  \BibitemOpen
  \bibfield  {author} {\bibinfo {author} {\bibfnamefont {J.}~\bibnamefont
  {Lee}}\ and\ \bibinfo {author} {\bibfnamefont {J.}~\bibnamefont
  {Kosterlitz}},\ }\href@noop {} {\bibfield  {journal} {\bibinfo  {journal}
  {Physical Review B}\ }\textbf {\bibinfo {volume} {43}},\ \bibinfo {pages}
  {3265} (\bibinfo {year} {1991})}\BibitemShut {NoStop}%
\bibitem [{\citenamefont {Binder}\ and\ \citenamefont
  {Landau}(1984)}]{Binder1984}%
  \BibitemOpen
  \bibfield  {author} {\bibinfo {author} {\bibfnamefont {K.}~\bibnamefont
  {Binder}}\ and\ \bibinfo {author} {\bibfnamefont {D.~P.}\ \bibnamefont
  {Landau}},\ }\href {\doibase 10.1103/PhysRevB.30.1477} {\bibfield  {journal}
  {\bibinfo  {journal} {Phys. Rev. B}\ }\textbf {\bibinfo {volume} {30}},\
  \bibinfo {pages} {1477} (\bibinfo {year} {1984})}\BibitemShut {NoStop}%
\bibitem [{\citenamefont {Binder}(1987)}]{Binder1987}%
  \BibitemOpen
  \bibfield  {author} {\bibinfo {author} {\bibfnamefont {K.}~\bibnamefont
  {Binder}},\ }\href@noop {} {\bibfield  {journal} {\bibinfo  {journal}
  {Reports on progress in physics}\ }\textbf {\bibinfo {volume} {50}},\
  \bibinfo {pages} {783} (\bibinfo {year} {1987})}\BibitemShut {NoStop}%
\bibitem [{\citenamefont {Landau}\ and\ \citenamefont
  {Binder}(2014)}]{Landau2014}%
  \BibitemOpen
  \bibfield  {author} {\bibinfo {author} {\bibfnamefont {D.~P.}\ \bibnamefont
  {Landau}}\ and\ \bibinfo {author} {\bibfnamefont {K.}~\bibnamefont
  {Binder}},\ }\href@noop {} {\emph {\bibinfo {title} {A guide to Monte Carlo
  simulations in statistical physics}}}\ (\bibinfo  {publisher} {Cambridge
  university press},\ \bibinfo {year} {2014})\BibitemShut {NoStop}%
\bibitem [{\citenamefont {Schnabel}\ \emph {et~al.}(2011)\citenamefont
  {Schnabel}, \citenamefont {Seaton}, \citenamefont {Landau},\ and\
  \citenamefont {Bachmann}}]{Schnabel2011}%
  \BibitemOpen
  \bibfield  {author} {\bibinfo {author} {\bibfnamefont {S.}~\bibnamefont
  {Schnabel}}, \bibinfo {author} {\bibfnamefont {D.~T.}\ \bibnamefont
  {Seaton}}, \bibinfo {author} {\bibfnamefont {D.~P.}\ \bibnamefont {Landau}},
  \ and\ \bibinfo {author} {\bibfnamefont {M.}~\bibnamefont {Bachmann}},\
  }\href@noop {} {\bibfield  {journal} {\bibinfo  {journal} {Physical Review
  E}\ }\textbf {\bibinfo {volume} {84}},\ \bibinfo {pages} {011127} (\bibinfo
  {year} {2011})}\BibitemShut {NoStop}%
\bibitem [{\citenamefont {Wang}\ and\ \citenamefont
  {Landau}(2001{\natexlab{a}})}]{Wang2001}%
  \BibitemOpen
  \bibfield  {author} {\bibinfo {author} {\bibfnamefont {F.}~\bibnamefont
  {Wang}}\ and\ \bibinfo {author} {\bibfnamefont {D.}~\bibnamefont {Landau}},\
  }\href@noop {} {\bibfield  {journal} {\bibinfo  {journal} {Physical Review
  E}\ }\textbf {\bibinfo {volume} {64}},\ \bibinfo {pages} {056101} (\bibinfo
  {year} {2001}{\natexlab{a}})}\BibitemShut {NoStop}%
\bibitem [{\citenamefont {Wang}\ and\ \citenamefont
  {Landau}(2001{\natexlab{b}})}]{Wang2001a}%
  \BibitemOpen
  \bibfield  {author} {\bibinfo {author} {\bibfnamefont {F.}~\bibnamefont
  {Wang}}\ and\ \bibinfo {author} {\bibfnamefont {D.}~\bibnamefont {Landau}},\
  }\href@noop {} {\bibfield  {journal} {\bibinfo  {journal} {Physical review
  letters}\ }\textbf {\bibinfo {volume} {86}},\ \bibinfo {pages} {2050}
  (\bibinfo {year} {2001}{\natexlab{b}})}\BibitemShut {NoStop}%
\bibitem [{\citenamefont {Caparica}(2014)}]{Caparica2014}%
  \BibitemOpen
  \bibfield  {author} {\bibinfo {author} {\bibfnamefont {A.~A.}\ \bibnamefont
  {Caparica}},\ }\href {\doibase 10.1103/PhysRevE.89.043301} {\bibfield
  {journal} {\bibinfo  {journal} {Phys. Rev. E}\ }\textbf {\bibinfo {volume}
  {89}},\ \bibinfo {pages} {043301} (\bibinfo {year} {2014})}\BibitemShut
  {NoStop}%
\bibitem [{\citenamefont {Caparica}\ and\ \citenamefont
  {Cunha{-}Netto}(2012)}]{Caparica2012a}%
  \BibitemOpen
  \bibfield  {author} {\bibinfo {author} {\bibfnamefont {A.~A.}\ \bibnamefont
  {Caparica}}\ and\ \bibinfo {author} {\bibfnamefont {A.~G.}\ \bibnamefont
  {Cunha{-}Netto}},\ }\href@noop {} {\bibfield  {journal} {\bibinfo  {journal}
  {Phys. Rev. E}\ }\textbf {\bibinfo {volume} {85}},\ \bibinfo {pages} {046702}
  (\bibinfo {year} {2012})}\BibitemShut {NoStop}%
\bibitem [{\citenamefont {Ferreira}\ and\ \citenamefont
  {Caparica}(2012)}]{Ferreira2012a}%
  \BibitemOpen
  \bibfield  {author} {\bibinfo {author} {\bibfnamefont {L.}~\bibnamefont
  {Ferreira}}\ and\ \bibinfo {author} {\bibfnamefont {A.}~\bibnamefont
  {Caparica}},\ }\href {\doibase 10.1142/S0129183112400128} {\bibfield
  {journal} {\bibinfo  {journal} {International Journal of Modern Physics C}\
  }\textbf {\bibinfo {volume} {23}},\ \bibinfo {pages} {1240012} (\bibinfo
  {year} {2012})}\BibitemShut {NoStop}%
\bibitem [{\citenamefont {Ferreira}\ \emph {et~al.}(2012)\citenamefont
  {Ferreira}, \citenamefont {Caparica}, \citenamefont {Neto},\ and\
  \citenamefont {Galiceanu}}]{Ferreira2012}%
  \BibitemOpen
  \bibfield  {author} {\bibinfo {author} {\bibfnamefont {L.~S.}\ \bibnamefont
  {Ferreira}}, \bibinfo {author} {\bibfnamefont {{\'A}.~A.}\ \bibnamefont
  {Caparica}}, \bibinfo {author} {\bibfnamefont {M.~A.}\ \bibnamefont {Neto}},
  \ and\ \bibinfo {author} {\bibfnamefont {M.~D.}\ \bibnamefont {Galiceanu}},\
  }\href {http://stacks.iop.org/1742-5468/2012/i=10/a=P10028} {\bibfield
  {journal} {\bibinfo  {journal} {Journal of Statistical Mechanics: Theory and
  Experiment}\ }\textbf {\bibinfo {volume} {2012}},\ \bibinfo {pages} {P10028}
  (\bibinfo {year} {2012})}\BibitemShut {NoStop}%
\bibitem [{\citenamefont {Post}\ \emph {et~al.}(2018)\citenamefont {Post},
  \citenamefont {McLeod}, \citenamefont {Hepting}, \citenamefont {Bluschke},
  \citenamefont {Wang}, \citenamefont {Cristiani}, \citenamefont {Logvenov},
  \citenamefont {Charnukha}, \citenamefont {Ni}, \citenamefont {Radhakrishnan},
  \citenamefont {Minola}, \citenamefont {Pasupathy}, \citenamefont {Boris},
  \citenamefont {Benckiser}, \citenamefont {Dahmen}, \citenamefont {Carlson},
  \citenamefont {Keimer},\ and\ \citenamefont {Basov}}]{Post2018}%
  \BibitemOpen
  \bibfield  {author} {\bibinfo {author} {\bibfnamefont {K.~W.}\ \bibnamefont
  {Post}}, \bibinfo {author} {\bibfnamefont {A.~S.}\ \bibnamefont {McLeod}},
  \bibinfo {author} {\bibfnamefont {M.}~\bibnamefont {Hepting}}, \bibinfo
  {author} {\bibfnamefont {M.}~\bibnamefont {Bluschke}}, \bibinfo {author}
  {\bibfnamefont {Y.}~\bibnamefont {Wang}}, \bibinfo {author} {\bibfnamefont
  {G.}~\bibnamefont {Cristiani}}, \bibinfo {author} {\bibfnamefont
  {G.}~\bibnamefont {Logvenov}}, \bibinfo {author} {\bibfnamefont
  {A.}~\bibnamefont {Charnukha}}, \bibinfo {author} {\bibfnamefont {G.~X.}\
  \bibnamefont {Ni}}, \bibinfo {author} {\bibfnamefont {P.}~\bibnamefont
  {Radhakrishnan}}, \bibinfo {author} {\bibfnamefont {M.}~\bibnamefont
  {Minola}}, \bibinfo {author} {\bibfnamefont {A.}~\bibnamefont {Pasupathy}},
  \bibinfo {author} {\bibfnamefont {A.~V.}\ \bibnamefont {Boris}}, \bibinfo
  {author} {\bibfnamefont {E.}~\bibnamefont {Benckiser}}, \bibinfo {author}
  {\bibfnamefont {K.~A.}\ \bibnamefont {Dahmen}}, \bibinfo {author}
  {\bibfnamefont {E.~W.}\ \bibnamefont {Carlson}}, \bibinfo {author}
  {\bibfnamefont {B.}~\bibnamefont {Keimer}}, \ and\ \bibinfo {author}
  {\bibfnamefont {D.~N.}\ \bibnamefont {Basov}},\ }\href {\doibase
  10.1038/s41567-018-0201-1} {\bibfield  {journal} {\bibinfo  {journal} {Nature
  Physics}\ } (\bibinfo {year} {2018}),\ 10.1038/s41567-018-0201-1}\BibitemShut
  {NoStop}%
\end{thebibliography}
\end{document}